\documentstyle[12pt]{article}
\begin{document}
 \bibliographystyle{unsrt}
 \vbox{\vspace{6mm}}

\begin{center}
{\Large  Quantum State Reconstruction of a 
Bose-Einstein Condensate} 
\end{center} 

\bigskip

\begin{center}
Stefano Mancini and Paolo Tombesi
\end{center}

\bigskip

\begin{center}
{\it Dipartimento di Matematica e Fisica,\\
Universit\'a di Camerino, I-62032 Camerino, Italy\\
and Istituto Nazionale di Fisica della Materia, Italy}
\end{center}

\bigskip
\bigskip
\bigskip

\begin{abstract}
We propose a tomographic scheme to reconstruct 
the quantum state of a Bose-Einstein condensate,
exploiting 
the radiation field as a probe and considering the
atomic internal degrees of freedom. The density 
matrix in the number state basis can be directly retrieved          
from the atom counting probabilities.
\end{abstract}

PACS numbers: 03.65.Bz, 03.75.Fi, 32.80.-t

\newpage

Since the Einstein's generalization of 
the black-body law derived by Bose,
where the prediction that identical atoms considered 
as an ideal gas at sufficiently low
temperature would become tight together in the lowest 
quantum state to form a condensate was made,
there was no experimental evidence of this phenomenon, 
meanly due to the extremely low temperature
at which it should occur. Finally, laser-based and 
magneto-optical trapping techniques were
developped to cool and trap atoms at the microkelvin 
level, preventing their solidification and/or
liquefaction. Then, in 1995 groups at JILA \cite{JILA} 
and MIT \cite{MIT} obtained the
experimental evidence for the condensation in dilute 
atomic gases: the Bose-Einstein condensation
(BEC). The BEC is a macroscopic occupation of the 
ground state of the gas and is one important
paradigm of quantum statistical mechanics. The density 
distribution of the condensate should be
represented by a macroscopic wave function with a well 
defined amplitude and phase, because of the
transition from disordered to coherent matter waves. 
Very recently high-contrast interference of 
two independent condensates were observed
\cite{Ketterle} showing that a condensate has a well 
defined phase.
Although the evidence of the macroscopic phase seems well
 established by the experimental results of
Ref. \cite{Ketterle} what is the BEC ground state is 
still an open question and one should
devise a technique able to measure it. 

In the last years in quantum optics the  state
 measurement of an electromagnetic field has
become a very studied subject, both from theoretical 
and experimentral points of view, after the
pioneering work by Vogel and Risken \cite{Risken}. 
The fundamental idea is that in repeated
experiments the outcomes of the homodyne measurement 
of the output light give the marginal
distribution of the Wigner function of the field. 
By varying the local oscillator phase one can
reconstruct the whole Wigner function, then the 
state of the field. This is not, however, the only
way to measure the quantum state of the radiation, 
because it was recently shown 
\cite{Wallentoviz,Banaszek,noi} that the reconstruction 
is possible 
by directly sampling the number of photons.
Since the atoms in BEC are bosons as well, one can 
think to apply the same argument to reconstruct
the condensate's state.
Recently, another approach to reconstruct the BEC 
density matrix
appeared in quant-phys 
files \cite{BTW}. The latter is however
based on the knowledge of the relative phase of two 
non interacting condensates which are different
for their hyperfine structures. 
It also needs of a beam splitter modeled for example by
an output coupler for atoms \cite{coupler}. 
Instead, we shall consider a reconstruction scheme
only based on atom counting, which exploits the interaction 
between the 
condensate atoms and a radiation field.

To this end we follow Javanainen \cite{juha94} and consider 
a Bose gas consisting of atoms 
moving in an isotropic harmonic oscillatory potential.
The eigenstates of the center of mass motion 
$|{\bf n}\rangle$ 
are labeled 
by the  vector index ${\bf n}=(n_x,n_y,n_z)$ 
of positive integers.
For the internal state of each atom we adopt the conventional 
two-state model:
ground state $|g\rangle$, excited state $|e\rangle$
which we assume having a long enough life-time $\tau_a$.
These are separated by the frequency $\omega_{e,g}$, and 
the dipole moment matrix between the states is ${\bf d}$. 
A radiation field, whose plane wave modes 
are enumerate by the index $q$, that incorporate both the wave 
vector ${\bf q}$ of the photon  and the polarization vector 
${\bf e}_q$, interacts with the ideal
gas. The frequency of the mode $q$ is $\Omega_q=|{\bf q}|$
(we shall use here the convention $\hbar=c=1$).

The Hamiltonian for the system consisting of 
the Bose gas and the photons thus is \cite{juha94}
\begin{eqnarray}\label{Htot}
H&=&\sum_{\bf n}\left[\epsilon_{\bf n}
b^{\dag}_{g,{\bf n}}b_{g,{\bf n}}
+(\epsilon_{\bf n}+\omega_{e,g})
b^{\dag}_{e,{\bf n}}b_{e,{\bf n}}\right]
+\sum_q\Omega_qa^{\dag}_qa_q\nonumber\\
&-&\sum_{{\bf n},{\bf n}',q}\left[
\xi(q)\langle{\bf n}'|e^{i{\bf q}{\cdot}{\bf r}}|
{\bf n}\rangle
b^{\dag}_{e,{\bf n}'}b_{g,{\bf n}}a_q
+{\rm H.c.}\right]\,.
\end{eqnarray}
The first two terms are the energies of the ground state and 
excited state atoms, the third term is the Hamiltonian for 
the free photon field and the final term governs the atom-field
interactions. 
For instance, a process 
in which absorption of  a photon in a state $q$ converts 
a ground-state atom in the c.m. state $|{\bf n}\rangle$ to 
an excited atom in the c.m. state $|{\bf n}'\rangle$, is 
governed by the matrix element $-\xi(q)\langle{\bf n}'|
e^{i{\bf q}{\cdot}{\bf r}}|{\bf n}\rangle$. The coupling 
coefficient pertaining to the internal states is 
$\xi(q)=\sqrt{\Omega_q/2V}\,{\bf e}_q\cdot{\bf d}$, where 
$V$ is the quantization volume. 
The explicit matrix elements in (\ref{Htot}) is the 
overlap between the c.m. state $|{\bf n}'\rangle$ 
and the c.m. state  $|{\bf n}\rangle$ shifted by the 
momentum ${\bf q}$.
The matrix element ${\bf d}$ is taken such that the 
correct optical 
linewidth $\gamma=d^2\omega_{e,g}^3/6\pi$ results
(for Cs atoms it is  $\gamma\approx (2\pi)\,2\; {\rm MHz}$).

In reality, for atoms moving inside a 
trap the total Hamiltonian should 
also contain 
terms representing the interaction between 
atoms, however, these 
can be neglected if we assume low enough density.
We further assume that the condensate is probed by 
a weak laser beam whose mode index and frequency are 
denoted by $k$ and $\Omega$.
The latter will be resonant with the trapped atoms,
i.e. $\Omega=\omega_{e,g}\approx 
(2\pi)\,4.0\times 10^{14}\;{\rm Hz}$ for Cs atoms.
As usual, the external field can be regarded as classical, 
and the corresponding photon operator can be treated as 
a $c$-number, i.e. we make the replacement $a_k\to\alpha_k$.
The spontaneous widths of the excited state are much larger
than any relevant c.m. frequency $\epsilon_{\bf n}$;
consequently we proceed from now on as if $\epsilon_{\bf n}=0$.
Then, the Hamiltonian (\ref{Htot})
will become
\begin{equation}
\label{Hint}
H=\omega_{e,g}\sum_{\bf n}
b^{\dag}_{e,{\bf n}}b_{e,{\bf n}}
-\sum_{{\bf n},{\bf n}'}\left[
\theta e^{-i\Omega t+i\phi}\langle{\bf n}'|
e^{i{\bf k}\cdot{\bf r}}|{\bf n}\rangle
b^{\dag}_{e,{\bf n}'}b_{g,{\bf n}}+{\rm H.c.}\right]\,,
\end{equation}
where $\theta=\xi(k)|\alpha_k|$ is the Rabi frequency 
that ensues when a classical field drives the internal 
transition in a single atom, and $\phi=\arg\alpha_k$.
 
Usually the condensate is modeled by 
treating its annihiliation operator as a {\it c}-number, 
i.e. $b_{g, {\bf 0}}\to \sqrt{N_c}$ with $N_c$ the number 
of atoms in the condensate.
Though this is true in the mean \cite{juha94,you94},
we continue to regard it as a purely quantum operator.
On the other hand, as it results from the first term inside 
the square 
brackets of Eq. (\ref{Hint}), the driving field only 
couples states in 
pairs $\{ |g\rangle\otimes |{\bf n}\rangle\,,\,
|e\rangle\otimes |\psi_{\bf n}\rangle\}$, where 
$|e\rangle\otimes |\psi_{\bf n}\rangle$ 
is obtained with the momentum translation 
${\bf k}$ from the excited state
$|e\rangle\otimes |{\bf n}\rangle$.
We now assume that (almost) all atoms are in the condensate,
therefore all matrix elements referring to the state 
$|g\rangle\otimes |{\bf 0}\rangle$ should pick up 
a large multiplier ($\approx\sqrt{N_c}$).
This allows us to retain in our theory only the condensate 
and its corresponding excited state  
$|e\rangle\otimes |\psi_{\bf 0}\rangle$ \cite{juha94}.
One then could resort to the boson operators
\begin{eqnarray}\label{bdef}
b_1&=&b_{g,{\bf 0}}\\
b^{\dag}_2&=&\sum_{{\bf n}'}
\langle{\bf n}'|e^{i{\bf k}\cdot{\bf r}}|0\rangle
b^{\dag}_{e,{\bf n}'}\,,
\end{eqnarray}
so that the mode 1 represents atoms in the internal 
ground state (and c.m. 
ground state), while mode 2 represents atoms in 
the internal excited states (no matter of what c.m. state).

In view of the above assumptions and  for
$\Omega=\omega_{e,g}$, 
Eq. (\ref{Hint})  
can be simply written as
\begin{equation}\label{H}
H=\Omega b^{\dag}_2b_2-\theta\left[
b_2^{\dag}b_1e^{-i\Omega t+i\phi}
+b_2b_1^{\dag}e^{i\Omega t-i\phi}\right]\,,
\end{equation}
where the phase $\phi$ is governed by the applied field.

If the atomic density operator 
at the initial observation time is $\rho$,
then its evolution, in a frame rotating at
the frequency of the applied field, 
will be given by
\begin{equation}\label{rhoev}
\rho(\tau)=
K(\Theta,\phi)^{\dag}\rho K(\Theta,\phi)
\end{equation}
with
\begin{equation}\label{KTh}
K(\Theta,\phi)=\exp\left[
i\Theta\left(b_1^{\dag}b_2e^{i\phi}
+b_1b_2^{\dag}e^{-i\phi}\right)\right]\,;\quad
\Theta=\theta \tau\,.
\end{equation}
Soon after this evolution, 
the system is left to expand ballistically 
and the number 
of atoms 
are counted, with some selective detectors able to discriminate 
atoms in the excited state from those in the ground state.
In order to simplify the presentation we assume that the 
measurement time is shorter than $\tau_a$, otherwise the 
detection efficiency should be considered as in \cite{leo,noi},
while for sake of simplicity we assume unity 
detection efficiency.

The evolution operator can be written in terms 
of generators of the $SU(2)$ group \cite{groups}
\begin{equation}\label{su2}
J_+=b^{\dag}_1b_2\,;\quad
J_-=b_1b^{\dag}_2\,;\quad
J_0=\frac{1}{2}\left(b_1^{\dag}b_1-b_2^{\dag}b_2\right)\,,
\end{equation}
obtaining 
\begin{equation}\label{Ksu2}
K(\Theta,\phi)=\exp\left[
i\Theta\left(J_+e^{i\phi}+J_-e^{-i\phi}\right)\right]\,.
\end{equation}

The probability of counting $n_1$ atoms in the mode 1 and $n_2$ 
in the mode 2 at the detector, 
for fixed $\phi$ and $\Theta$, is
\begin{eqnarray}\label{P}
P(n_1,n_2,\Theta,\phi)&=&\langle n_1,n_2|K(\Theta,\phi)^{\dag}
\rho K(\Theta,\phi)|n_1,n_2\rangle \nonumber\\
&=&\sum_{l_1,m_1}\sum_{l_2,m_2}
{\cal R}_{l_1,m_1;l_2,m_2}^{n_1,n_2}(\Theta,\phi)
\rho_{l_1,m_1;l_2,m_2}\,,
\end{eqnarray}
where
\begin{equation}\label{Tbig}
{\cal R}^{n_1,n_2}_{l_1,m_1;l_2,m_2}(\Theta,\phi)
=\langle l_1,l_2|K(\Theta,\phi)|n_1,n_2\rangle^*\times
\langle m_1,m_2|K(\Theta,\phi)|n_1,n_2\rangle\,.
\end{equation}
The rotation matrix elements in the number 
state basis are obtained by 
using the Baker-Campbell-Hausdorff formula for the Lie 
algebra of $SU(2)$ \cite{BCH}
\begin{eqnarray}\label{Kn1n2}
\langle m_1,m_2|K(\Theta,\phi)|n_1,n_2\rangle&=&
\sum_j^{n_1}
{\sum_k}'
\sqrt{\frac{m_1!(m_1+k)!n_1!(n_2+j)!}
{(m_1-k)!m_2!(n_1-j)!n_2!}}\\
&\times&\frac{(i\tanh\Theta)^
{j+k}}{j!k!}
\exp\left[-\ln\{\cosh\Theta\}(n_1-n_2-2j)\right]
e^{i\phi(k-j)}\nonumber\,,
\end{eqnarray}
where the prime sign on the sum means that $k$ assumes 
the values $n_2-m_2+j$ and
$m_1-n_1+j$.

Eqs. (\ref{Tbig}) and (\ref{Kn1n2}) show that ${\cal R}$ 
is simply 
a linear operator connecting the density matrix elements 
to the number
probability data which gives rise
to an overdetermined system of linear equations
provided one knows the probability as a function of the 
experimentally controlled parameters $\phi$ and 
$\Theta$. 
Hence, it can be inverted \cite{Robi}
to get the density matrix 
elements in the number basis by just collecting the 
experimental data for various values of $\phi$ and
$\Theta$. 
Since we do not assume any fixed total number of atoms,
a truncation procedure should be employed.
Let us label with $N_1$ and $N_2$ the upper bounds for 
the number of atoms 
in the modes 1 and 2 respectively, then $\rho$
should be intended as a complex vector with 
$(N_1+1)^2\times(N_2+1)^2$ 
elements and $P$ as a real vector with 
$(N_1+1)\times(N_2+1)\times N_{\phi}\times N_{\Theta}$ 
elements, 
where $N_{\phi}$ and $N_{\Theta}$ are the total 
number of values
of  $\phi$ and $\Theta$ used.
Provided to have $N_{\phi}\ge(N_1+1)$ 
and $N_{\Theta}\ge(N_2+1)$,
or viceversa, one can invert Eq. (\ref{P}) 
to get the density matrix 
from the measured probabilities
by using some numerical methods \cite{Robi}.
Essentially, by defining the hermitian matrix
$G={\cal R}^{\dag}{\cal R}$, we can compute
\begin{equation}\label{inv}
\rho=G^{-1}{\cal R}^{\dag}P\,,
\end{equation}
provided to have $G$ non singular.

It is worth noting that the above introduced density 
matrix does not 
concern only the condensate atoms in the ground state, 
but also those in 
the excited states.
Since, due to the low density assumption,
 we are neglecting the
collisional effects between the two 
modes, it is reasonable to write it in a factorized 
form before the radiation field 
action, i.e.
\begin{equation}\label{rhofac}
\rho=\rho^{(1)}\otimes\rho^{(2)}\,,
\end{equation}
where now $\rho^{(1)}$ represents the density operator 
for the condensate.
Of course once one has calculated Eq. (\ref{inv}), the 
matrix elements
$\rho^{(1)}_{l_1,m_1}$ immediately follow by tracing out 
the mode 2.

The above procedure, however, although correct in principle, 
seems rather 
difficult to implement numerically due to the very high 
dimensions of the 
involved matrices. To simplify the scheme it is possible, 
with a good 
approximation, to assume that in the initial state all 
atoms are in their 
internal ground state, 
i.e. $\rho^{(2)}=|0\rangle_2\,{}_2\langle 0|$, then 
Eq. (\ref{P})
becomes
\begin{equation}\label{Pcond}
P(n_1,n_2,\Theta,\phi)=\sum_{l_1,m_1}
{\cal R}^{n_1,n_2}_{l_1,m_1;0,0}
(\Theta,\phi)\rho^{(1)}_{l_1,m_1}\,.
\end{equation}
In this case $\rho^{(1)}$ should be intended 
as a complex vector 
with $(N_1+1)^2$ elements,then Eq. (\ref{Pcond}) can be 
inverted by only 
varying the phase $\phi$ of the applied 
field, mantaining 
$\Theta$ fixed, for example. The phase of the external
field can be arbitrarily fixed with the initial measurement
because the reconstruction procedure only needs to specify
a given number of phases in the interval $[0,2\pi]$ starting
from any phase value.

In conlusion, we have presented an approach to the quantum 
state reconstruction of a BEC based on a  
tomographic procedure which seems easy to 
implement because one only needs to master 
the probing field and to measure the number 
probabilities of atomic 
detection.
It is worth noting that the presented method does not 
need the knowledge of the exact number of the trapped 
number of atoms but only an upper bound for it.
Of course, for a fixed  and known number of atoms 
the procedure is
simpler since one needs to only vary the phase of 
the  field independently
of the assumption on $\rho^{(2)}$.
Only in this case, however, one can get rid of non 
unity efficiency
of the detector, because by counting both species
of atoms one can 
discard those data which do not give the initial 
total number \cite{BTW}.
It is also worth noting that the presented 
procedure is well suited for 
condensates with a relatively small number 
of atoms, because of the 
number of data one should collect. This should not be
a limitation because if the condensate were in an almost
coherent state, as it is usually assumed, its wave function
should not depend too much on the number of atoms 
in the condensate.
Thus, BEC with a small number of atoms (say 50-100) 
should be obtained
in order to easily implement the present model, 
and in such a case the 
assumption we introduced might be better justified.
The realization of such a small condensate could
be the object of next generation experiments, once the
existence of a given phase for the condensate will be
better tested by other experimental groups. In our 
opinion the possibility of "measuring" the true
density matrix of the condensate is accessible and
worth considering.

\end{document}